\documentstyle[12pt]{article}

\def\br{\begin{thebibliography}{abc}}
\def\er{\end{thebibliography}}

\newcommand{\sect}[1]{\setcounter{equation}{0}\section{#1}}

\begin{document}

\begin{titlepage}

{\hspace*{\fill} SU-4240-602\\
\hspace*{\fill} hep-th/9503091 \\
\hspace*{\fill} February 1995\\ }

\bigskip\bigskip

\begin{center}
      {\Large\bf Chern-Simons terms in Noncommutative Geometry
and its application to Bilayer Quantum Hall Systems\\}
\end{center}

\bigskip

\begin{center}
{\large Varghese John,  Nguyen Ai Viet
\footnote{\noindent
On leave from High Energy Division, Centre of Theoretical Physics, P.O.Box 429
Bo Ho 10000, Hanoi, Vietnam.}
and Kameshwar C. Wali }
\hskip 3mm
\medskip\medskip \\
{\it Physics Department, Syracuse University\\
Syracuse , NY 13244-1130, USA.
{}~\footnote{ E-mail: aiviet@suhep.phy.syr.edu,
john@suhep.phy.syr.edu wali@suhep.phy.syr.edu}\\}
\end{center}

\bigskip
\begin{abstract}

 Considering bilayer systems as extensions of the planar ones by
an internal space of two discrete points, we use the  ideas of
Noncommutative Geometry to construct the gauge theories for these systems.
After integrating over the discrete space we find an effective
$2+1$ action involving an extra complex scalar field, which can be interpreted
as arising from the tunneling between the layers. The gauge fields are found in
different phases corresponding to the different correlations due to the
Coulomb interaction between the layers. In a particular phase, when the radial
part of the complex scalar field is a constant, we recover the Wen-Zee model
\cite{WEZE} of Bilayer Quantum Hall systems. There are some circumstances,
where this radial part may become dynamical and cause dissipation in the
oscillating supercurrent between the layers.

\end{abstract}

\end{titlepage}
\sect{Introduction}
The geometry of a bilayer system can be considered as the extension of the
$2+1$ dimensional space-time manifold {\cal M }by an internal discrete space of
two elements indexing the layers. The purpose of this paper is to explore
the possibility of applying the formalism of Noncommutative Geometry (NCG)
proposed recently by Connes \cite {Co1} to the study of such multi-layer
systems.

As a particular application of NCG, the Connes-Lott version of the Standard
Model \cite{CoLo} begins with two
copies of abstract space-time. Particles with different chiralities are
assumed to
exist on two different abstract sheets. The connection between the
sheets is
mediated by the Higgs fields that trigger spontaneous symmetry breaking to
give mass to the gauge fields. This formalism has also been applied to
two sheeted structures, not necessarily of different chiralities, to construct
a discretized version of Kaluza-Klein theories \cite{LVW}. In these examples,
although NCG has provided new tools and new insights, its claim to be a
harbinger of a generalized quantum theory is not immediately apparent.
The physical models that have emerged are essentially classical. We shall see
that the physical bilayer systems that have attracted a great deal of attention
in the recent literature \cite{WEZE,ZEE,Fradkin}, are eminently suitable for
exploiting the NCG formalism. The quantum features of such systems are
naturally incorporated within the formalism. In addition, studies with such
realistic systems and consequences that can be experimentally tested provide
a fertile ground for understanding the physical basis of NCG.

One of the main results of our investigation is that within the framework of
NCG, the customary Maxwell-Chern-Simons terms do arise. Further, when the
formalism is specialized to describe the Bilayer Quantum Hall Systems, we
obtain a model containing two gauge fields and a complex scalar field that
appears naturally as a component along the discrete direction. In the special
case when the radial part of this complex field is assumed to be constant, we
recover the model of Wen and Zee of Bilayer Quantum Hall systems with tunneling
between the layers \cite{WEZE}. Thus the angular part of our scalar field
receives a physical
interpretation. It arises from the quantum tunneling processes between
the layers. In the Wen-Zee model a dissipationless
oscillating supercurrent is predicted. In our model, if the radial part of the
scalar field is position- and time- dependent, the supercurrent will dissipate.
Moreover, we can expect that such a general situation does arise in the
Wen-Zee model when one goes beyond the dilute monopole gas approximation and
the fugacity to create a monopole pair becomes position dependent.

The formalism of NCG also shows that there are two naturally occurring phases
in
the system considered. That such phases ( called in- and out-phases) arise from
the inter-layer Coulomb interaction has been demonstrated in
Ref.\cite{Fradkin}. We find
that the Bilayer Quantum Hall System is in general in a mixed phase, which
is a
superposition of the in- and out-phases. This mixed phase corresponds to the
physical situation when one combination of the gauge fields does not have a
topological mass but acquires a mass due to tunneling between the layers.

\sect{Noncommutative Geometric Formulation of Gauge theories
in Bilayer Quantum Hall Systems}
Following Connes and Lott \cite{CoLo} we  invoke the idea of NCG
to study the correlations between two layers. Although this model has been
applied to  the abstract ``{layers}" of right- and left-handed
particles in the
Standard Model, it is straightforward to translate it to the case of two
realistic layers. We use a formalism \cite{LVW}
which is more
transparent to the physicists.

Let us denote the electronic wave functions in the bilayer system by
$\psi ~ \in {\cal H} = {\cal H}_{1}  \oplus {\cal H}_{2}$

\begin{equation}
\psi = \pmatrix{\psi_{1} \cr
                \psi_{2} \cr}
\end{equation}
where ${\cal H}_I~~ (I= 1,2 )$ are the Hilbert spaces associated to the
electrons on the I'th layer.

The algebra of smooth functions
 ${\cal C}^\infty ({\cal M})$ on the manifold ${\cal M}$ is generalized to
${\cal A} ~=~ {\cal C}^\infty({\cal M})\oplus{\cal C}^\infty({\cal M})$ and any
element  $F \in {\cal A}$ can be written as
\begin{equation}
F  ~=~f_+(x)\pmatrix{1&0\cr
       0&1\cr} ~ + ~ f_-(x) \pmatrix{1&0\cr
                                   0&-1\cr} ~=~\pmatrix{f_1(x)&0\cr
                                                0&f_2(x)\cr}~,
\end{equation}
where $ f_{\pm}(x) = 1/2 (f_2(x) \pm f_1(x)) $.

The exterior derivative $ d= dx^\mu \partial_\mu~ (\mu = 0,1,2) $ is
generalized to
\begin{equation}
D = d.{\bf 1} + Q = D X^{M} D_M = DX^{\mu} D_{\mu} + DX^3 \sigma^\dagger D_3
\end{equation}
where $ M = 0,1,2,3 $ , ${\bf 1}$ is the $2 \times 2 $ unit matrix and
\begin{eqnarray} \label{DER}
 DX^\mu ~=~
\pmatrix{dx^\mu & 0 \cr
         0 & dx^\mu \cr} &~ ,~&
D_\mu ~=~ \pmatrix{\partial_{\mu} & 0\cr
                    0 & \partial_{\mu}\cr}, \cr
DX^3 ~=~ \pmatrix{0&\theta \cr
                  \theta &0\cr} &~,~&
D_3 ~=~ \pmatrix{~0 & m\cr
               - m & 0\cr} , \cr
 D_3 F ~=~ \left [ D_3, F \right ]~=~
\pmatrix {0& 2 m f_-(x)\cr
          2 m f_-(x)& 0\cr} &~,~&
 \sigma ~=~\pmatrix {~ 0 & 1\cr
                    -1 & 0 \cr},
\end{eqnarray}
m is a parameter with dimension of mass and $\theta$ is an anti-hermitian
Clifford operator.

Here $D_3 $ is a derivative in the following sense:

1) $ D_3 F$ resembles the derivative
$\delta F \over \delta l $ where $\delta l$ is the distance between the two
layers $ \left ( m^{-1} = \delta l , \delta F = f_2 - f_1 \right) $ .

2) It satisfies the Newton-Leibnitz rule
$D_3 (F G ) = D_3 F. G + F . D_3 G $ .

Acting on functions the exterior derivative gives
\begin{eqnarray}\label{DER0}
DF\doteq (~DX^\mu D_\mu ~+~ DX^3\sigma^\dagger D_3  )F =
                            \pmatrix{df_1& \theta m(f_2-f_1)\cr
                             \theta m(f_1-f_2) &df_2\cr} ,
\end{eqnarray}
which is obviously hermitian.
The general hermitian 1-forms are given by the following matrices:
\begin{eqnarray}\label{ONEFORM}
 A ~=~ \pmatrix{  a_1           & \theta \phi^* \cr
             -\theta \phi   &    a_2 \cr} & ~ = ~& DX^{\mu} A_{\mu}
+ DX^3 A_3 ,\nonumber \\
A_\mu ~=~ \pmatrix{ a_{1\mu} & 0 \cr
                       0 & a_{2\mu} \cr } &~,~&
A_3 ~=~ \pmatrix{ - \phi & 0 \cr
                   ~  0   & \phi^* \cr } ,
\end{eqnarray}
where $ a_I =dx^\mu a_{I\mu} $ is an 1-form on the I'th layer.
In the generalized 1-form in
Eq.(\ref{ONEFORM}), besides the ordinary gauge fields on the two layers we
also find a new complex scalar.

To define the field strength 2-form and the higher forms, we need a
definition of the wedge product. It turns out that for a description of
Bilayer Quantum Hall
systems the proper wedge product must be chosen as  \footnote{ Let us mention
that, an alternate definition of the wedge product has been used in \cite{LVW}
for gravity in NCG}:
\begin{eqnarray}\label{WEDGE}
DX^\mu \wedge DX^\nu & =  &
   -DX^\nu \wedge DX^\mu ,\nonumber\\
DX^3 \wedge DX^\mu & =  &
   -DX^\mu \wedge DX^3 , \nonumber\\
DX^3 \wedge DX^3 & =  & {\bf 1}.
\end{eqnarray}

Traditionally, the field strength 2-form $ \Omega $ can be defined as follows
\begin{equation}\label{TWOFORM1}
 \Omega =DA + A \wedge A \doteq DX^M \wedge DX^N \Omega _{MN} .
\end{equation}
The field strength defined in Eq.(\ref{TWOFORM1}) has the geometric meaning of
curvature. The model constructed from this field strength contains
two gauge fields coupled to the scalar field with a quartic potential. This
scalar field breaks the gauge symmetry $ U(1)\times U(1)$ {\it spontaneously}.
However, in Bilayer Quantum Hall
systems with tunneling effects, symmetries are {\it explicitly} broken.
So, we are looking for an alternate way.
 As the gauge fields are abelian, the field
strength 2-form $ \Omega $ can also be generalized as follows

\begin{equation}\label{TWOFORM2}
 \Omega =DA   \doteq DX^M \wedge DX^N \Omega _{MN}  .
\end{equation}
The components $ \Omega _{MN}$ in this case are given by:
\begin{eqnarray}
\Omega _{\mu \nu} & =& {1\over 2} \left( \partial_{\mu} A_{\nu}
-\partial _{\nu} A_{\nu}\right ) =  {1\over 2}\pmatrix{ f_{1 \mu \nu} & 0 \cr
                                   0 & f_{2\mu \nu} },  \cr  \nonumber\\
\Omega _{\mu 3} & =& {1\over 2} \left( \partial _{\mu} A_{3}
-\sigma^{\dagger}D_{3} A_{\mu}\right ) = {1\over 2}
\pmatrix{- \partial_{\mu} \phi +2m a_{-\mu} & 0 \cr
                                   0 &\partial_{\mu} \phi^* - 2m a_{-\mu}
 } , \cr  \nonumber\\
\Omega _{3 3 } &=& -m \left( \phi + \phi^*\right) \pmatrix{1&0\cr
                                   0&-1\cr} .
\end{eqnarray}
 As we
shall see, although its geometric meaning is not obvious, the field
strength defined in Eq.(\ref{TWOFORM2}) is relevant for a description of
Bilayer Quantum Hall Systems.
So, hereafter we will
use the field strength defined in Eq.(\ref{TWOFORM2}) to build a generalized
Chern-Simons-Maxwell gauge model of Bilayer Quantum Hall Systems.
Our choice of using this field strength
is motivated by two considerations: i) As a direct generalization, it includes
all terms of the ordinary Chern-Simons-Maxwell gauge theory, ii) Nature
requires
that such systems have explicitly broken symmetries.

\vskip 0.5cm
{\large \it 2.1 The Chern-Simons terms :}

The ordinary Chern-Simons terms $\int d^3 x \varepsilon^{\mu \nu
\lambda}a_{\mu}
\partial_{\nu} a_{\lambda} $ can be found in the following generalized
Chern-Pontryagin term
\begin{equation}\label{CS}
{\cal L} _{CP} =  {1\over m} Tr \int d^3 x ~K ~~
\varepsilon ^{MNPQ}~ \Omega_{MN} ~\Omega_{PQ} + h.c ,
\end {equation}
where K is a generalized function used to define the measure,
 ${1\over m} Tr= \int d x^3 $
is the discrete analogue of the integration over the internal
space.

The Chern-Pontryagin term (\ref{CS}) then becomes:
\begin{eqnarray}\label{CSa}
{\cal L}_{CP} & = &
   {1\over m} Tr \int d^3 x ~K ~~
\varepsilon ^{\mu \nu \lambda} 2m a_{-\mu} \left(D_{\nu} A_{\lambda}
-D_{\lambda} A_{\nu} \right)\nonumber\\
&-&
{1\over m}Tr \int d^3 x ~K ~~
\varepsilon ^{\mu \nu \lambda} D_{\mu} A_3 \left(D_{\nu} A_{\lambda}
-D_{\lambda} A_{\nu} \right) + h.c \nonumber\\
&= & \int d^3 x
\varepsilon ^{\mu \nu \lambda} 2 a_{-\mu} \left(-k_1\partial_{\nu} a_{1\lambda}
+k_2 \partial_{\nu} a_{2\lambda} \right) \nonumber\\
&-&  {1\over m} \int d^3 x
\varepsilon ^{\mu \nu \lambda}  \left(k_1\partial_{\mu }\phi
\partial_{\nu} a_{1\lambda}
-k_2 \partial_{\mu} \phi^*\partial_{\nu} a_{2\lambda} \right) + h.c .
\end{eqnarray}

The vector $A_\mu  $ is related to the physical gauge fields
$\alpha_{I\mu}(x)$ as follows
\begin{equation}\label{gen}
A_\mu =  \pmatrix{g_1 & 0 \cr
              0 & g_2 \cr} \pmatrix{\alpha_{1\mu} & 0 \cr
                                     0 & \alpha_{2\mu} \cr}
= G \pmatrix{\alpha_{1\mu } & 0 \cr
                0 & \alpha_{2\mu } \cr},
\end{equation}
where $ g_I~ (I=1,2)$ is the coupling constant on the I'th layer and $G$ is the
the matrix of the coupling constants
\begin{equation}\label{COUPLING}
G ~~=~~\pmatrix{ g_1 & 0 \cr
                  0 & g_2 \cr} .
\end{equation}

The first term in Eq.(\ref{CSa}) can be rearranged in the standard form:
$
\int d^3 x \kappa _{IJ}
\varepsilon ^{\mu \nu \lambda}  \alpha_{I\mu}
\partial_{\nu} \alpha_{J\lambda}
$, where the matrix $\kappa =2 \pmatrix{ +k_1 g_1 ^2 & - k_1 g_1 g_2 \cr
            - k_2 g_1 g_2 & k_2 g_ 2 ^2\cr }$.

As shown by Halperin \cite{Halperin}, the matrix $ \kappa $ is of the
symmetric form

\begin{equation}
\kappa ={1\over 4\pi} \pmatrix{ l & n \cr
            n & j\cr }
\end{equation}
 where l, j, n are integers.
Comparing the matrix $\kappa$ of our model with this form, we are lead to
the condition $k_1 = k_2 = k$ in order that $ \kappa $ be symmetric. Then
\begin{equation}
\kappa =2 k \pmatrix{ g_1^2 & -g_1 g_2 \cr
            - g_1 g_2 & g_2 ^2\cr }
\end{equation}
and if we require that the coupling constants
in the two layers  are the same in absolute value $|g_1|=
|g_2| $ , we have two
different cases:

Case (1): $g_1 = -g_2 = g$
\begin{eqnarray}\label{INPHASE}
\kappa_{IJ} & = & 2k g^2\pmatrix{ 1 & 1 \cr
            1 & 1\cr } , \nonumber \\
A_{\mu}& = & g \pmatrix{\alpha _{1 \mu} & 0 \cr
                      0 & -\alpha_{2 \mu}\cr } .
\end{eqnarray}

Case (2): $g_1 = g_2 = g$

\begin{eqnarray}\label{OUTPHASE}
\kappa_{IJ} & = &
      2k g^2\pmatrix{ ~1 & - 1 \cr
            -1 &~ 1\cr }, \nonumber \\
A_{\mu} & = & g \pmatrix{ \alpha_{1 \mu}& 0 \cr
                           0   &  \alpha_{2 \mu} \cr }.
\end{eqnarray}

These two cases correspond to two phases of Bilayer Quantum Hall systems:
the in and out phases discussed in Ref.\cite{Fradkin}. Comparing the matrix
$\kappa$ in two above cases with the standard form of the Chern-Simons terms
of Bilayer Quantum Hall systems \cite{WEZE,ZEE} we have
$ k= {j\over 8\pi g^2}$, where $j$ is an integer.

Without violating the physical condition that the coupling constants in
different layers should have the same absolute value, we may consider
the general case in Eq.(\ref{gen}) as a mixing
of the phases (\ref{INPHASE}) and (\ref{OUTPHASE}):
\begin{eqnarray}\label{explain}
A_{\mu} &=& g \left( cos {\beta \over 2}
\pmatrix{ \alpha_{1 \mu} & o \cr
           0 & \alpha_{2 \mu} \cr}
+ sin{\beta \over 2} \pmatrix{ \alpha_{1 \mu} & o \cr
           0 & -\alpha_{2 \mu} \cr}\right)\nonumber\\
&=&\pmatrix{ g(cos{\beta\over 2}+ sin {\beta \over 2})\alpha_{1 \mu} & o \cr
           0 & g(cos{\beta\over 2}- sin {\beta \over 2})\alpha_{2 \mu} \cr},
\end{eqnarray}
where $0\leq \beta \leq {\pi}$ is the mixing angle of the two
phases.  The coupling constant matrix $ G$ in Eq.(\ref{COUPLING})
now becomes
\begin{equation}
G = \pmatrix{g_1 & 0 \cr
             0 & g_2 \cr} =
g \pmatrix { cos{\beta\over 2} + sin{\beta \over 2} & 0 \cr
                  0 & cos{\beta \over 2} - sin{\beta \over 2} \cr}
\end{equation}
The mixing angles $ \beta = {\pi \over 2}, {3\pi \over 2} $ are automatically
excluded
because then one of the matrix elements in Eq.(\ref{explain}) is zero
and  the corresponding Chern-Simons term on one layer vanishes.

In this paper we will discuss this general case and then specialize to the
particular cases by setting $\beta = \pi $ (the in phase) and $\beta = 0 $
(the out phase) when it becomes necessary.

In the general case the $\kappa $ - matrix is:
\begin{equation}\label{QUANT}
  {1\over 4\pi}\pmatrix{ l & n  \cr
           n & j \cr}
= 2kg^2 \pmatrix{ 1+sin \beta &-cos \beta  \cr
          -cos \beta & 1-sin \beta \cr}
\end{equation}

As $ det \kappa = 0 $ one eigenvalue of the matrix $\kappa $ is always zero,
while the other is not.
This in turn means that there are two linear combinations of the gauge
fields one of which is topologically massive due to the Chern-Simons term
and the other is massless. Diagonalizing the matrix $ \kappa $ we find these
combinations are:
\begin{equation}
 \alpha^\beta_{\pm \lambda}~=~ cos{\beta \over 2}
\alpha_{+ \lambda} \pm sin{\beta \over 2} \alpha_{- \lambda}.
\end{equation}
where $ \alpha_{\pm \lambda} = \alpha_2 \pm \alpha_1 $.

The Eq.(\ref{QUANT}) imposes a  strong restriction on the mixing angle
$\beta $ as well as on the filling factor.
 This condition and its physical interpretation will be discussed elsewhere
\cite{JOVI2}.

The second term in Eq. (\ref{CSa}) turns out to be:
\begin{equation}
{2kg\over m} \int d^3x \varepsilon^{\mu \nu \lambda}
\partial_{\mu} (\phi +\phi^*) \partial_\nu \alpha^\beta_{+\lambda} .
\end{equation}
If $ \phi + \phi^* $ is smooth this term is
a surface term and can be neglected.

So far we have obtained a model that
represents a Bilayer Quantum Hall system without tunneling between two layers.
In our NCG model one combination of gauge fields, $\alpha^{\beta} _{-\mu}$
is massless.  The physical meaning of the massless mode is the following:
When two layers are close enough
the Coulomb interaction between  them is strong.
It is possible to have a correlated fluctuation of
the densities of the two layers which costs no energy. This corresponds to
the massless mode. In phase and out phase correspond to the charge density
correlations between charges of the same sign and the opposite sign
respectively.

\vskip 0.5cm
{\large \it 2.2 The Maxwell term :}

We now consider the effect of adding a Maxwell term to this analysis.
As in the ordinary Maxwell theory, here  we also need a metric structure:
\begin{eqnarray}
< DX^{\mu} , DX^{\nu} >  &=  & g^{\mu \nu}{\bf 1} \nonumber\\
< DX^{\mu} , DX^{3} >  &=  & 0 \nonumber\\
< DX^{3} , DX^{3} >  &=  & {\bf 1} \nonumber\\
< DX^{\mu}\wedge DX^{\nu} ,DX^{\rho}\wedge DX^{\sigma}> & =&
{1\over 2} \left( g^{\mu \sigma} g^{\nu \rho} - g^{\mu \rho} g^{ \nu
\sigma}\right) {\bf 1}\nonumber\\
< DX^{\mu}\wedge DX^{3} , DX^{\nu}\wedge DX^{3}> & =& {1\over 2}
g^{\mu \nu}{\bf 1} \nonumber\\
< DX^{3}\wedge DX^{3} ,DX^{3}\wedge DX^{3}> & =&{\bf 1}
\end{eqnarray}

The Maxwell action is defined as a direct generalization of the ordinary
Maxwell term $ ( 1/g^2) F^2 $
\begin{equation}\label{MX}
{\cal L} _{Maxwell} =  {1\over m} Tr \int d^3 x~~ G^{-2}~~
 <\Omega ^2>.
\end{equation}

It is straightforward to calculate the Maxwell term (\ref{MX}) with the
definition of the field strength given in Eq.(\ref{TWOFORM2}). We find
\begin{eqnarray}
{\cal L}_{Maxwell} & = & \int d^3 x -{1\over 4} f_1^2 -{1\over 4} f_2^2
+ {1 \over g^2 cos^2 \beta }
\partial_{\mu} \phi^* \partial^{\mu} \phi  + {2m \over g}{\partial ^{\mu}
(\phi^* +\phi)\over cos^2 \beta} \alpha^{\beta}_{+ \mu}\cr
& + & {4m^2 \over cos^2 \beta} \left(
\alpha^\beta_{+ \mu} \right)^2 + {2m^2\over
g^2 cos^2 \beta}
\left( \phi + \phi^*\right)^2
\end{eqnarray}
where $f_{I \mu \nu}= \partial_{\mu} \alpha_{I\nu} -\partial_{\nu}
\alpha_{I\mu}$ is the gauge field strength on the I'th layer.

Let us represent the complex field $\phi$ as $\varphi(x) e^{i\theta (x) \over
2}$ and write down the full theory as follows:
\begin{eqnarray}
{\cal L} &=& {\cal L}_{CP} + {\cal L} _{Maxwell}~=~
\int d^3 x \varepsilon^{\mu \nu \lambda}
\kappa_{IJ} \alpha_{\mu I} \partial_ { \nu} \alpha_{J \lambda}\cr
&+& {4kg \over m} \int d^3 x \varepsilon^{\mu \nu \lambda}
\partial_{\mu} ( \varphi(x) cos {\theta(x) \over 2})
\partial_\nu \alpha^\beta_{+\lambda }\cr
&+& \int d^3 x  -{1\over4} (f^\beta_{-\mu \nu})^2 - {1\over 4}
( f^\beta_{+\mu \nu})^2
+{\varphi^2\over 4g^2 cos^2\beta} (\partial_{\mu} \theta)^2
+ {1\over g^2 cos^2 \beta} (\partial_\mu \varphi)^2 \cr
&+&{4m^2\over cos^2\beta}\int d^3 x \left( \alpha^\beta_{+\mu} \right)^2
 + {2\varphi^2\over g^2 } cos ^2 {\theta(x)\over 2}
+{1\over mg} \partial^{\mu}( \varphi cos {\theta \over 2})
\alpha^{\beta}_{+\mu} .
\end{eqnarray}
For a moment let us assume that the
dynamical field $\varphi(x)$ is frozen to some constant value
$\varphi_{0}$ then we have:
\begin{eqnarray}
{\cal L} &=&
\int d^3 x \varepsilon^{\mu \nu \lambda}
\kappa_{IJ} \alpha_{\mu I} \partial_ { \nu} \alpha_{J \lambda}
+ {4kg\varphi_0 \over m } \int d^3 x \varepsilon^{\mu \nu \lambda}
\partial_{\mu} (  cos {\theta(x) \over 2}) \partial_\nu \alpha^\beta_{+\lambda}
\cr
&+& \int d^3 x-{1\over4} (f^\beta_{-\mu \nu})^2 - {1\over 4} (f^\beta_{+\mu
\nu})^2
+{\varphi_0 ^2\over 4g^2 cos^2 \beta } (\partial_{\mu} \theta)^2
+{8m^2\varphi_0^2 \over g^2 cos^2 \beta } cos^2{\theta(x) \over 2}  \cr
&+&\int d^3x  {4m^2 \over cos^2 \beta } (\alpha^\beta_{+\mu}) ^2
 + 2mg \varphi_0 \partial^\mu (cos{\theta \over 2}) \alpha^\beta_{+\mu} .
\end{eqnarray}

In order to compare with the
results of Wen and Zee  \cite{WEZE} we specialize to the case of in-phase
and obtain :
\begin{eqnarray}\label{THIRTYONE}
& {\cal L} &= {\cal L}_1 + {\cal L}_2 + {\cal L}_3 \nonumber\\
& {\rm where } &   \nonumber  \\
& {\cal L }_1 & =
\int d^3 x \varepsilon^{\mu \nu \lambda}
\kappa_{IJ} \alpha_{\mu I} \partial_ { \nu} \alpha_{J \lambda}
-{1\over4} f_+^2 - {1\over 4} f_-^2 \nonumber \\
& {\cal L}_2 & =
\int d^3 x {\varphi_0 ^2\over 4g^2 } (\partial_{\mu} \theta)^2
+\zeta cos\theta(x)
\cr
& {\cal L}_3 & = \int d^3 x
{4kg\varphi_0 \over m}\varepsilon^{\mu \nu \lambda}
\partial_{\mu} (  cos {\theta(x) \over 2})
 \partial_{\nu} \alpha_{+ \lambda} + {4m^2 } \alpha_{+\mu } ^2  \cr
& &
+2mg \varphi_0 \partial^\mu( cos{\theta(x) \over 2}) \alpha_{+\mu} .
\end{eqnarray}

 Eq.(\ref{THIRTYONE}) expresses the results of our model in the in-phase case
when the
radial part $\varphi(x)$ of the scalar field is assumed to be constant. We note
that the Maxwell-Chern-Simons terms that appear in $ {\cal L}_1 $
describe the Hall fluids in the two layers
in the absence of tunneling. Tunneling processes involving the transition of an
electron from one layer to another leads to the
non-conservation of the current $J_{-\mu} = j_{2\mu}-j_{1\mu} $. As discussed
in Ref.\cite{WEZE}, the consequent non-vanishing divergence of this current can
be effectively described as a monopole. When tunneling occurs
in large numbers, the effective monopole configuration can be described by an
order parameter, that characterizes the monopole gas.
Polyakov has shown \cite {POLY} that a weakly interacting monopole gas can
be represented by a scalar field $\theta(x)$ with a sine-Gordon interaction,
exactly as in the $ {\cal L}_2 $ in Eq.(\ref{THIRTYONE}).

Hence the terms ${\cal L}_1$ and ${\cal L}_2$ in Eq.(\ref{THIRTYONE}) recover
the
Wen-Zee model of
Bilayer Quantum Hall Systems with the quantum tunneling between the two layers.
The tunneling process is
described by the order parameter $\theta(x)$ which is the angular part of our
scalar field $\phi(x)$.

If $\theta(x)$ is a non-singular field, the first term in ${\cal L}_3 $
can be written as a surface term and does not
contribute to the dynamics.

The second term in ${\cal L}_3 $ is an explicit mass
term for gauge field $\alpha_{+\mu}$. Together with the Chern-Simons term it
gives rise to two non-vanishing poles
for the propagator of the gauge field $ \alpha_{+\mu}$\cite{PIRA}.

The last term represents a coupling between the fields $ \alpha_{+\mu}$ and $
\theta (x)$. That is to say, the Quantum Hall Fluid described by the gauge
field $\alpha_{+\mu}(x)$ affects the tunneling processes between two layers in
our
model. This is a new feature of our model that has not been discussed in
\cite{WEZE}. On general grounds, there is no reason against the presence of
such an interaction.

Now we discuss the origin of the field $\varphi(x) $ and the
situation when $\varphi(x)$ becomes dynamical.
It seems natural to expect that monopoles are described by
a complex scalar field as they carry a non-trivial charge.
Only the angular part of this field has been used in the
Coulomb gas description in literature.
However, we can make heuristic arguments ( a complete discussion is beyond the
scope of this paper) to show that the radial field may be relevant in actual
physical situations.

By coupling the complex scalar field $\phi $ to an external electromagnetic
field through the covariant derivative of the scalar field $\phi(x)$ in
Eq.(2.27), we can find the electromagnetic current to be
\begin{equation}
J^\mu_{em}= i( \partial^\mu\phi^*(x) \phi(x) - \phi^*(x) \partial^\mu \phi(x)
= 2 \varphi^2(x)\partial^\mu \theta(x).
\end{equation}
Hence, the charge density $ \rho = 2 \varphi^2(x)\partial^0 \theta(x)$
is proportional to $\varphi^2(x)$ and the time derivative of $\theta(x)$.

The physical meaning of $ \varphi^2(x)$ can be read from the following
expression of the fugacity
\begin{equation}
\zeta = {4m^2 \varphi^2(x) \over g^2},
\end{equation}
which is a consequence of our model.
Once the field $\varphi$ is no longer
frozen to a constant, the fugacity acquires a position and time dependence.
This means that the probability to create a monopole will depend on position
and time.

In the classical analysis of Polyakov \cite{POLY} the interaction of
monopoles was assumed to be weak, and the fluctuations in the density of
monopoles could be neglected. So the fugacity could be considered constant.
But this is not the most general situation as position and time dependent
interactions between charges and monopoles should affect the probability to
create a monopole.
In Bilayer Quantum Hall Systems the position dependence
could also originate from the fact that there are inhomogeneities in
the system. This can arise from impurities in
the bilayer samples or from polarization effects
when a voltage is applied to the edges of the samples.
Wen and Zee have also observed that an edge effect would lead to
a position dependent fugacity. From our picture, we can conclude that the
field $\varphi (x)$ may be important near the edges due to the
presence of inhomogeneities. So the field $\varphi(x)$ is an additional
order parameter to describe the inhomogeneity of the tunneling processes
in Bilayer Quantum Hall Systems.

In the case of constant fugacity, as a consequence of the potential
$cos \theta(x) $, Wen and Zee have been able to derive a dissipationless
oscillation between the two layers, whose frequency is determined by a dc
voltage applied across the two layers. This current resembles the Josephson
current in the superconductor-insulator-superconductor junction.
However, by the arguments given above we believe that there are actual
physical situations where the field $\varphi(x)$ is dynamical. In that case,
the
field $\varphi(x)$ will cause dissipation in the oscillating
periodic current. This dissipation may explain the difficulty in observing the
oscillating current predicted by Wen and Zee.

\sect{Conclusion}
We now summarize the results of our paper and state our conclusions.
In this paper we have formulated the Chern-Simons-Maxwell theory for  Bilayer
Quantum Hall Systems using a NCG approach \cite{LVW}\footnote{ At this
point we would like to mention that an earlier attempt was made by Bellisard,
who has applied NCG to the single-layer Quantum Hall system \cite{Bellisard}.}.
The ordinary Chern-Simons terms, which describe the long-distance physics of
the Quantum Hall fluid cannot be found in a generalized Chern-Simons term
\cite{FROH},
but rather in a generalized Chern-Pontryagin term. This term
leads to a model, where one combination of the gauge fields remains massless
corresponding to the strongly correlated fluctuation of the electrons on
different
layers. Besides the in phase and out phase, we find that
the NCG construction allows a general phase, which can be considered
as mixing of these two phases.
In our generalized Maxwell term the complex scalar field in
the hermitian gauge connection 1-form becomes dynamical and
describes the tunneling
between the layers. As a particular case of our construction, the in
phase case leads to a model which is very similar to the one used by Wen and
Zee
\cite{WEZE} to study tunneling effects in Bilayer Quantum Hall systems.
Additionally, we also find a new interaction term between the fields
$ \alpha_{+\mu}(x)$ and $ \theta(x)$ and an explicit mass term of
$\alpha_{+\mu}(x)$.

At this point we can conclude that the NCG formalism we have
presented
gives us a coherent description of the physics of bilayer systems.
The physical model that we have studied demonstrates that in NCG the
scalar part of the gauge fields in fact has its origin in some quantum
processes like the tunneling between the two layers. Moreover, the formalism
gives rise to some new features whose consequences need further study

\vskip 0.5cm

{\large \it  Acknowledgments:}

One of us (V.J) would like to acknowledge useful discussions
with A.P. Balachandran, L. Chandar, G.
Jungman, A. Momen, B. Sathiapalan and S. Vaidya.
N.A.V thanks Profs. Quang Ho-Kim and Chia-Hsiung Tze for inspiring
conversations.

\end{document}